\documentclass[11pt,a4paper]{article}
\usepackage{latexsym,epsfig}
\usepackage{amsmath,amssymb,amsthm}
\usepackage{fullpage}
\usepackage{graphicx,caption}
\usepackage[textsize=footnotesize]{todonotes}
\usepackage{epstopdf}
\usepackage{color}
\usepackage{url,hyperref}

\usepackage{graphicx}
\usepackage{psfrag}
\usepackage{multicol}
\usepackage{enumerate}
\usepackage[affil-it]{authblk}
\newcommand{\myL}{{hooked L-shape}}

\usepackage{subcaption}

\graphicspath{{.},{figures/}}

\newtheorem{theorem}{Theorem}
\newtheorem{lemma}{Lemma}[section]

\newtheorem{definition}[lemma]{Definition}

\newtheorem{corollary}[theorem]{Corollary}

\begin{document}

\title{4-connected planar graphs are in $B_3$-EPG}

\author[a]{Therese Biedl}
\author[b]{Claire Pennarun}

\renewcommand\Affilfont{\itshape\small}
\affil[a]{David R. Cheriton School of Computer Science, University of Waterloo, {biedl@cs.uwaterloo.ca}}
\affil[b]{LaBRI, Univ. Bordeaux, UMR 5800, F-33400 Talence, \par claire.pennarun@labri.fr}

\date{}

\maketitle


\begin{abstract}
We show that every 4-connected planar graph has a $B_3$-EPG representation, i.e., every vertex is represented by a curve on the grid with at most three bends, and two vertices are adjacent if and only if the corresponding curves share an edge of the grid.
Our construction is based on a modification of the representation by touching thickened $L$-shapes proposed by Gon{\c c}alves et al.~\cite{GIP17}.
\end{abstract}

\section{Introduction}
\label{sec:Intro}

A \emph{VPG-representation} of a graph $G$ consists of assigning to each vertex of $G$ a path on a rectangular grid such that two vertices are adjacent in $G$ if and only if their corresponding paths share at least one point.  It is called a {\em $B_k$-VPG-representation} if every vertex-path has at most $k$ bends.  Very recently, it was shown that every planar graph has a $B_1$-VPG representation, and in fact, every vertex-curve has the shape of an $L$~\cite{GIP17}.

\begin{figure}[htbp!]
\centering
\includegraphics[page=2,scale=0.7,trim=0 20 0 20,clip]{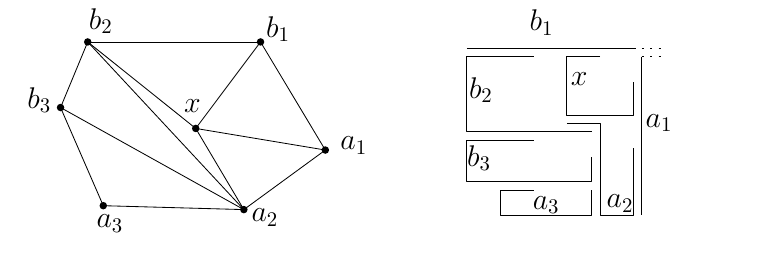}
\caption{A 2-sided near-triangulation $T$ and a $B_3$-EPG representation of
$T-(a_1,b_1)$ created with our algorithm. (To represent $(a_1,b_1)$, add the dotted 
segments.)} 
\label{fig:example}
\end{figure}

An \emph{EPG-representation} of a graph $G$ also consists of assigning to each vertex of $G$ a path in the rectangular grid, but this time two vertices are adjacent in $G$ if and only if their corresponding paths share at least one grid edge. It is called a {\em $B_k$-EPG-representation} if every vertex-path has at most $k$ bends (see Figure~\ref{fig:example} for an example).
It is known that every planar graph has a $B_4$-EPG-representation,
and that there are planar graphs that need at least three bends for every vertex-path \cite{HKU14}.

We refer to \cite{GIP17} and \cite{HKU14} for more background and results
regarding VPG-representations and EPG-representations.
The contribution of this short note is the following:

\begin{theorem} \label{th:4-conn}
All 4-connected planar triangulations have a $B_3$-EPG-representation.
\end{theorem}

In consequence, any graph $G$ that is an induced subgraph of a 
4-connected planar trangulation $H$ also has a $B_3$-EPG-representation, by
taking the one of $H$ and deleting the paths of vertices in $H-G$.   Using
this, we can show:

\begin{corollary} \label{cor:noSepTriangle}
Let $G$ be a planar graph with an embedding without separating triangle.
Then $G$ has a $B_3$-EPG-representation.
\end{corollary}

We strongly suspect that separating triangles
can also be handled, i.e., any planar graph has a $B_3$-EPG-representation,
but this remains for future work.

\section{Proof of Theorem~\ref{th:4-conn}}

\begin{figure}[htbp]
\hspace*{\fill}
\begin{subfigure}[b]{0.25\linewidth}
\includegraphics[page=1,width=0.99\linewidth]{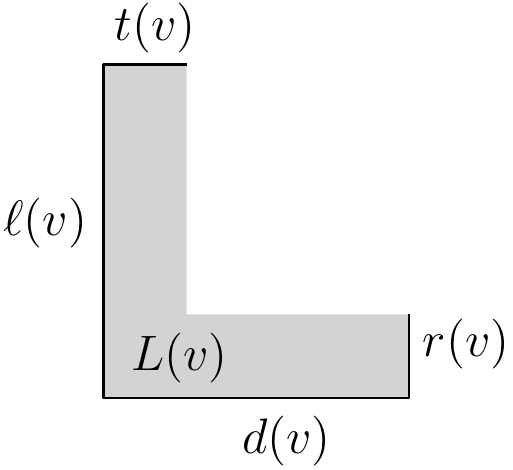}
\caption{}
\end{subfigure}
\hspace*{\fill}
\begin{subfigure}[b]{0.25\linewidth}
\includegraphics[page=2,width=0.99\linewidth]{hook.pdf}
\caption{}
\end{subfigure}
\hspace*{\fill}
\caption{Converting a thickened L-shape $L(v)$ into a \myL in two different ways.}
\label{fig:hook}
\end{figure}

We first give an idea for the proof of Theorem~\ref{th:4-conn} that seems simple,
but does not quite work out.
Gon\c{c}alves et al.~\cite{GIP17} showed that every 4-connected
planar graph can be represented as the contact graph of thickened
unrotated L-shapes.  Using this, one can define a 
grid-path $P(v)$ for every  vertex $v$
by walking along the outside edges of the L-shape $L(v)$ 
representing $v$ (see Figure~\ref{fig:hook}(a)).
Clearly then $P(v)$ has three bends. 

However, the details of this approach are not so easy.  Let us assume that
some L-shape $L(w)$ attached on the vertical edge incident to the reflex
vertex of $L(v)$.  Since the curve $P(v)$
runs along the outside edges of $L(v)$, then $P(w)$ does not share a 
grid-edge with $L(v)$.  One could fix this by letting $P(v)$ run along
the inside edges of $L(v)$ instead (see Figure~\ref{fig:hook}(b)), 
but then the same issue arises with
L-shapes attaching on the left side of $L(v)$.  One can argue that with
a suitable choice (and by extending the ``hooks'' if needed) one can always
obtain a $B_3$-EPG representation this way,
but the construction rules are quite complicated and require deep
insights into how the L-shape contact representation of \cite{GIP17} was
obtained.  
Rather than explaining these details, we instead create the
$B_3$-EPG contact representation from scratch, with the same recursive 
approach that Gon\c{c}alves et al.~used.  

\subsection{Some definitions}

We review some definitions from \cite{GIP17} first.
Let $G$ be a planar graph with a fixed planar embedding.  We say that
$G$ is a {\em near-triangulation} if it is a triangulated disk, i.e.,
it is 2-connected and every inner face is a triangle, and if $G$ has
no {\em separating triangle}, i.e., a triangle $T$ 
that is not a face.%
\footnote{Some other references (see \cite{CG09,CGO10}) use the term {\em W-triangulation} for such graphs, since they were first studied by Whitney \cite{Whit31}.} 

\begin{definition}\cite{GIP17}  A near-triangulation $G$ {\em is 
2-sided} if we can enumerate the outer-face of $G$ in 
clockwise order as $a_1,\dots,a_p,b_q,b_{q-1},\dots,b_2,b_1$ such that
there is no edge $(a_i,a_j)$ for $|j-i|>1$ and no edge $(b_i,b_j)$ 
for $|j-i|>1$.
\end{definition}

\begin{figure}[htbp!]
\centering
\includegraphics[scale=0.7]{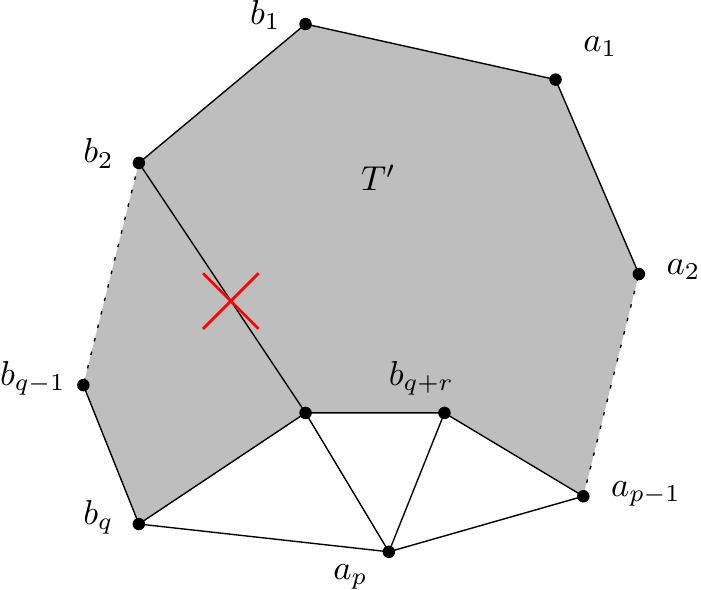}~~~~~~
\includegraphics[scale=0.7]{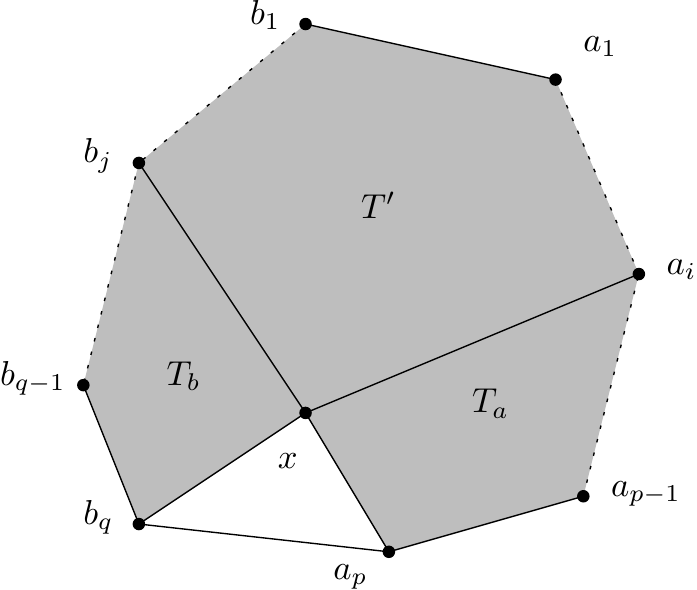}
\caption{The $a_p$-removal operation and the split operation on a 2-sided near-triangulation (adapted from~\cite{GIP17} with permission).
}
\label{fig:operations}
\end{figure}

\begin{lemma}\cite{GIP17}  
\label{lem:split}
Let $T$ be a 2-sided near-triangulation with
at least 4 vertices.  Then one of the following situations occurs (see Figure~\ref{fig:operations}):
\begin{itemize}
\item An {\em $a_p$-removal}:  We have $p\geq 2$, and if we remove vertex $a_p$,
	then the resulting graph $T'$ is a 2-sided near-triangulation with
	respect to the outer-face $a_1,\dots,a_{p-1},b_{q+r},b_{q+r-1},
	\dots,b_{q+1},\allowbreak b_q,\dots,b_1$, where $b_{q+1},\dots,b_{q+r}$ are
	the neighbours of $a_p$ not previously on the outer-face.
\item A {\em $b_q$-removal}:  We have $q\geq 2$, and if we remove vertex $b_q$,
	then the resulting graph $T'$ is a 2-sided near-triangulation with
	respect to the outer-face $a_1,\dots,a_p, a_{p+1},\dots,a_{p+r},
	b_{p-1},\dots,\allowbreak b_1$, where $a_{p+1},\dots,a_{p+r}$ are
	the neighbours of $b_q$ not previously on the outer-face.
\item A {\em split}:  The unique common neighbour $x$ of $a_p$ and $b_q$
	has a neighbour $a_i$ for $i<p$ and a neighbour $b_j$ for $j<q$, and with
	a suitable choice of such $i,j$ all of the following hold:
	\begin{itemize}
	\item The graph $T'$ bounded by $a_1,\dots,a_i,x,b_j,b_{j-1},\dots,b_1$
	is a 2-sided near-triangulation if we consider $x$ to be the new
	$a_{i+1}$ for this graph.
	\item The graph $T_a$ bounded by $a_i,a_{i+1},\dots,a_p,x$ 
	is a 2-sided near-triangulation if we consider $x$ to be the new
	$b_{1}$ for this graph.
	\item The graph $T_b$ bounded by $x,b_q,b_{q-1},\dots,b_j$ 
	is a 2-sided near-triangulation if we consider $x$ to be the new
	$a_{1}$ for this graph.
	\end{itemize}
\end{itemize}
\end{lemma}

\subsection{The invariant}

Observe that any 4-connected triangulation is a 2-sided near-triangulation
if we enumerate the outer-face arbitrarily as $a_1,a_2,b_1$.  To prove
Theorem~\ref{th:4-conn}, it therefore suffices to create 
$B_3$-EPG-representations for any 2-sided near-triangulation $T$.  
We will actually
create a $B_3$-VPG representation of $T-(a_1,b_1)$, but with such restrictions
on the curves for $a_1$ and $b_1$ that adding a shared grid-edge for $P(a_1)$
and $P(b_1)$ can be done easily by adding short horizontal segments (see
Figure~\ref{fig:example}).
For purpose of illustration, we consider that two paths share a grid-edge if they are drawn very close to each other on parallel lines. 

Our representation is not only a $B_3$-EPG-representation, but the grid
path $P(v)$ representing vertex $v$ has a particular shape (up to
lengthening of segments).  Define
a {\em \myL} to be the following (see also Figure~\ref{fig:hook}):  
It consists of a vertical (``left'') segment $\ell(v)$ and a horizontal (``down'')
segment $d(v)$ that form an $L$, i.e., the bottom end of $\ell(v)$ coincides with
the left end of $d(v)$.  It may or may not have two more segments $t(v)$ (``top'')
or $r(v)$ (``right''), where one end of $t(v)$ coincides with the top end of $\ell(v)$
and one end of $r(v)$ coincides with the right end of $d(v)$.

\begin{figure}[htbp!]
\hspace*{\fill}
\includegraphics[scale=0.8,page=1]{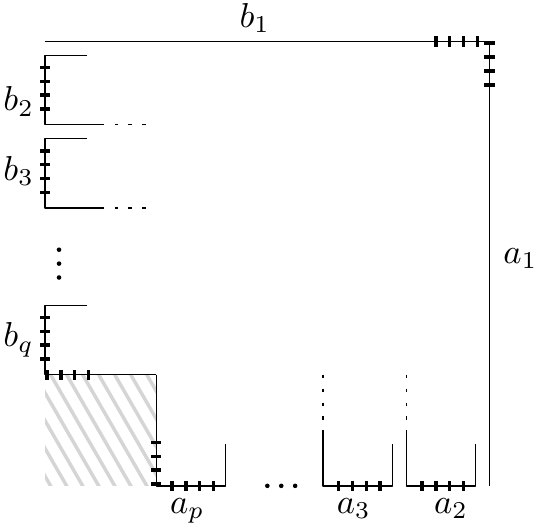}
\hspace*{\fill}
\includegraphics[scale=0.8,page=2]{invariant.pdf}
\hspace*{\fill}
\caption{The invariant of the bounding box $B$ of the EPG-representation of $G - (a_1,b_1)$ with outer-face $a_1,\dots,a_p,b_q,\dots,b_1$. The dashed region is empty.  
Hatched grid-lines are exclusively owned. 
The right figure indicates the special case $q=1$.}
\label{fig:invariant}
\end{figure}

We show the following result:

\begin{lemma}
Let $G$ be a 2-sided near-triangulation with outer-face $a_1,\dots,a_p,b_q,\dots,b_1$.  Then $G-(a_1,b_1)$ has a $B_3$-EPG-representation $\Gamma$ 
that satisfies the following (see Figure~\ref{fig:invariant}):
\begin{enumerate}
\item $P(a_1)$ is a vertical segment $\ell(a_1)$,
\item $P(b_1)$ is a horizontal segment $d(b_1)$, 
\item all other grid-paths $P(v)$ are {\myL}s,
\item \label{it:boundary}
	the boundary of the representation is a (possibly degenerate) 
	orthogonal 6-gon $B$ with the following properties:
	\begin{enumerate}
	\item The open south-west quadrant of the reflex corner of $B$ is empty.
	\item $d(b_1)$ forms the top side of $B$.  The rightmost
		grid-edge of $d(b_1)$ belongs exclusively to $d(b_1)$ (i.e.,
		belongs to no other grid-path).
	\item $\ell(a_1)$ forms the right side of $B$.  The topmost
		grid-edge of $\ell(a_1)$ belongs exclusively to $\ell(a_1)$.
	\item \label{it:left} The left side of $B$ contains from top to bottom:
		the left endpoint of $d(b_1)$, grid-edges that belong 
		exclusively to $\ell(b_2)$, grid-edges that belong exclusively 
		to $\ell(b_3)$, $\dots$, grid-edges that belong exclusively 
		to $\ell(b_q)$.
	\item \label{it:bottom} The bottom side of $B$ contains from right to left:
		the bottom endpoint of $\ell(b_1)$, grid-edges that belong 
		exclusively to $d(a_2)$, grid-edges that belong exclusively 
		to $d(a_3)$, $\dots$, grid-edges that belong exclusively to 
		$d(a_p)$.
	\item \label{it:ap} The vertical side incident to the reflex corner of $B$ belongs 
		to $\ell(a_p)$;
		the bottommost grid-edge of it belongs exclusively to $\ell(a_p)$.
	\item \label{it:bq} The horizontal side incident to the reflex corner of $B$ belongs 
		to $d(b_q)$;
		the leftmost grid-edge of it belongs exclusively to $d(b_q)$.
	\end{enumerate}	
\item \label{it:tap} If $q=1$, then removing the segment $t(a_p)$ from the
representation gives an EPG-representation of $G-(a_1,b_1)-(a_p,b_1)$.  
In other words, with the exception of edge $(a_p,b_1)$, all incident edges 
of $a_p$ are realized at segments of $P(a_p)$ other than $t(a_p)$.
\item \label{it:rbq} If $p=1$, then removing the segment $r(b_q)$ from the
representation gives an EPG-representation of $G-(a_1,b_1)-(a_1,b_q)$.  
\end{enumerate}
\end{lemma}

\subsection{Constructing the representation}

\begin{figure}[htbp!]
\centering
\includegraphics[scale=0.9]{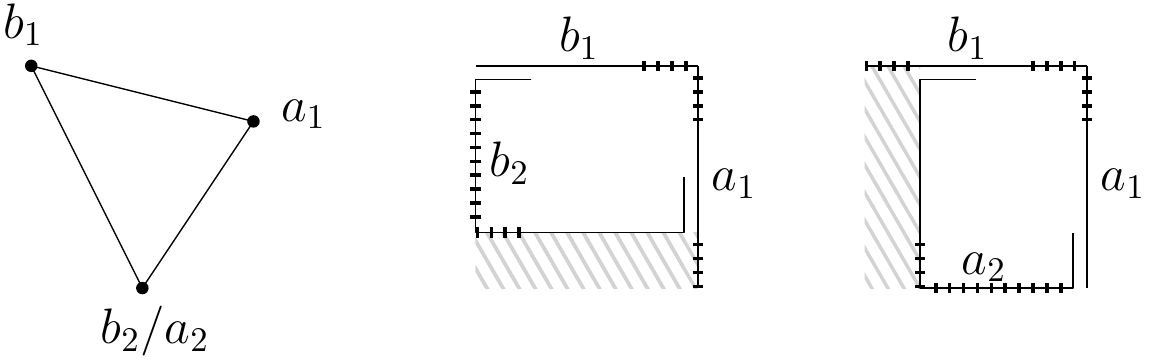}
\caption{The base case of the induction: $T$ is a triangle that can be represented using at most 3 bends per vertex-path. Recall that we represent $T-(a_1,b_1)$, hence the grid-paths of $a_1$ and $b_1$ have no edge in common.}
\label{fig:base-case}
\end{figure}

We prove the lemma by induction on the number of vertices.  In the
base case, $n=3$ and $T$ is a triangle.  Figure~\ref{fig:base-case} shows the construction,
both for when we enumerate the outer-face as $a_1,a_2,b_1$ and when we
enumerate it as $a_1,b_2,b_1$.  One easily verifies the claim.

Now presume that $n\geq 4$, and apply Lemma~\ref{lem:split} to split
$T$ into smaller 2-sided near-triangulation(s) $T'$ and (possibly) $T_a$
and $T_b$.  Recursively find $B_3$-EPG-representations for these subgraphs
(minus the edge  $(a_1,b_1)$).  It remains to argue how, for each
operation, we combine the drawings and add grid-paths for missing vertices
and edges.

\begin{figure}[htbp!]
\hspace*{\fill}
\includegraphics[scale=0.75,page=1]{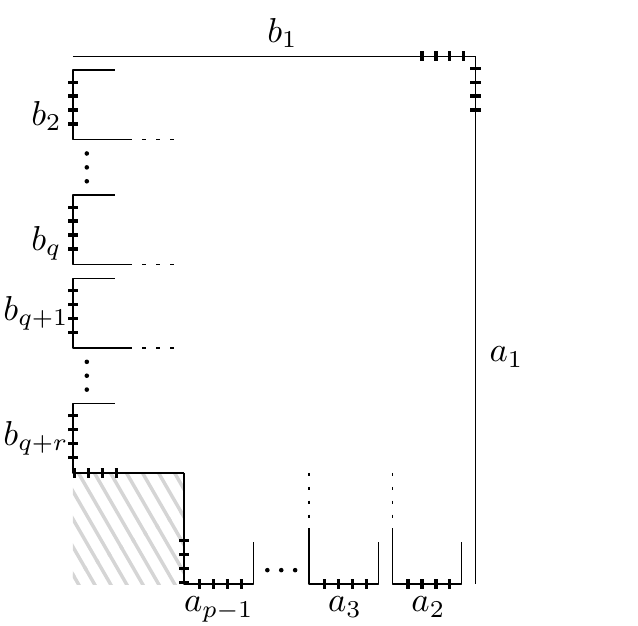}
\hspace*{\fill}
\includegraphics[scale=0.75,page=2]{ap-removal.pdf}
\hspace*{\fill}
\includegraphics[scale=0.75,page=3]{ap-removal.pdf}
\hspace*{\fill}
\caption{The $a_p$-removal operation: (a) The drawing of $T' - (a_1,b_1)$. (b) Extending the grid-paths $P(b_1),\dots,P(b_{q+r})$ leftwards and the grid-paths $P(a_1),\dots,P(a_{p-1})$ downwards allows to add a new grid-path for vertex $a_p$ (drawn heavier).  (c) The $a_p$-removal operation in the case $p=1$.}
\label{fig:ap-removal}
\end{figure}

\paragraph{$a_p$-removal}  Let $\Gamma'$ be the $B_3$-EPG-representation
obtained recursively for $T'-(a_1,b_1)$, and let $B'$ be its boundary
(see Figure~\ref{fig:ap-removal}).  Along the left side of $B'$, we 
encounter (from top to bottom) the 
left end of $d(b_1)$ and then $\ell(b_2),
\dots,\ell(b_{q+r})$.    We now change the grid-paths $P(b_1),\dots,P(b_{q+r})$
by moving the ends of their left sides leftward and extending the incident
segments correspondingly.  By Invariant~\ref{it:left},  
we know that no grid-edge of
$\ell(b_j)$ (for $2\leq j\leq q$) belongs to any other grid-path, so
moving $\ell(b_j)$ leftward does not change which grid-paths 
intersect.  We move $P(b_1),\dots,P(b_q)$ leftward by two units, and
$P(b_{q+1}),\dots,P(b_{q+r})$ leftward by one unit. 

We now add a grid-path $P(a_p)$ for $a_p$ as follows.  Start at the
left endpoint $s$ of $d(a_{p-1})$ (or, for $p=2$, at the bottom endpoint 
of $\ell(a_1)$).    We know that the grid-edge above $s$ belongs exclusively
to $P(a_{p-1})$ by Invariant~\ref{it:ap}; add this grid-edge to $P(a_p)$.  
Now $P(a_p)$ and $P(a_{p-1})$ share a grid-edge as required.
Grid-path $P(a_p)$ continues from $s$ leftward until one unit before the
left side, and then goes upward until it hits $d(b_q)$, say at point $t$.  
(This must happen since we extended $d(b_q)$ leftward by two units.)  
Since we moved $\ell(b_{q+1}),\dots,\ell(b_{q+r})$ by one unit, the 
segment of $P(a_p)$
below $t$ shares grid-edges with $P(b_{q+1}),\dots,P(b_{q+r})$ as required.
Finally, add the grid-edge to the right of $t$ to $P(a_p)$; this
grid-edge is then shared by $a_p$ and $b_q$ (and also $b_{q+1}$ if
$r\geq 1$), and is not shared by any other grid-path since we moved
$\ell(b_q)$ and $\ell(b_{q+1})$ leftward.
One easily verifies that all conditions are satisfied. 

Figure~\ref{fig:ap-removal} also demonstrates the special
case $q=1$; observe that Invariant~\ref{it:tap} 
holds since $t(a_p)$ belongs to
$d(b_1)$ and $t(b_2)$ and to no other grid-path, 
but edge $(b_2,a_p)$ is also represented by
a common grid-edge of $\ell(a_p)$ and $\ell(b_2)$.

\paragraph{$b_q$-removal}    This case is symmetric to the case of
an $a_p$-removal and makes use of Invariants~\ref{it:bottom} and \ref{it:bq}.

\begin{figure}[htbp!]
\centering
\includegraphics[scale=0.9]{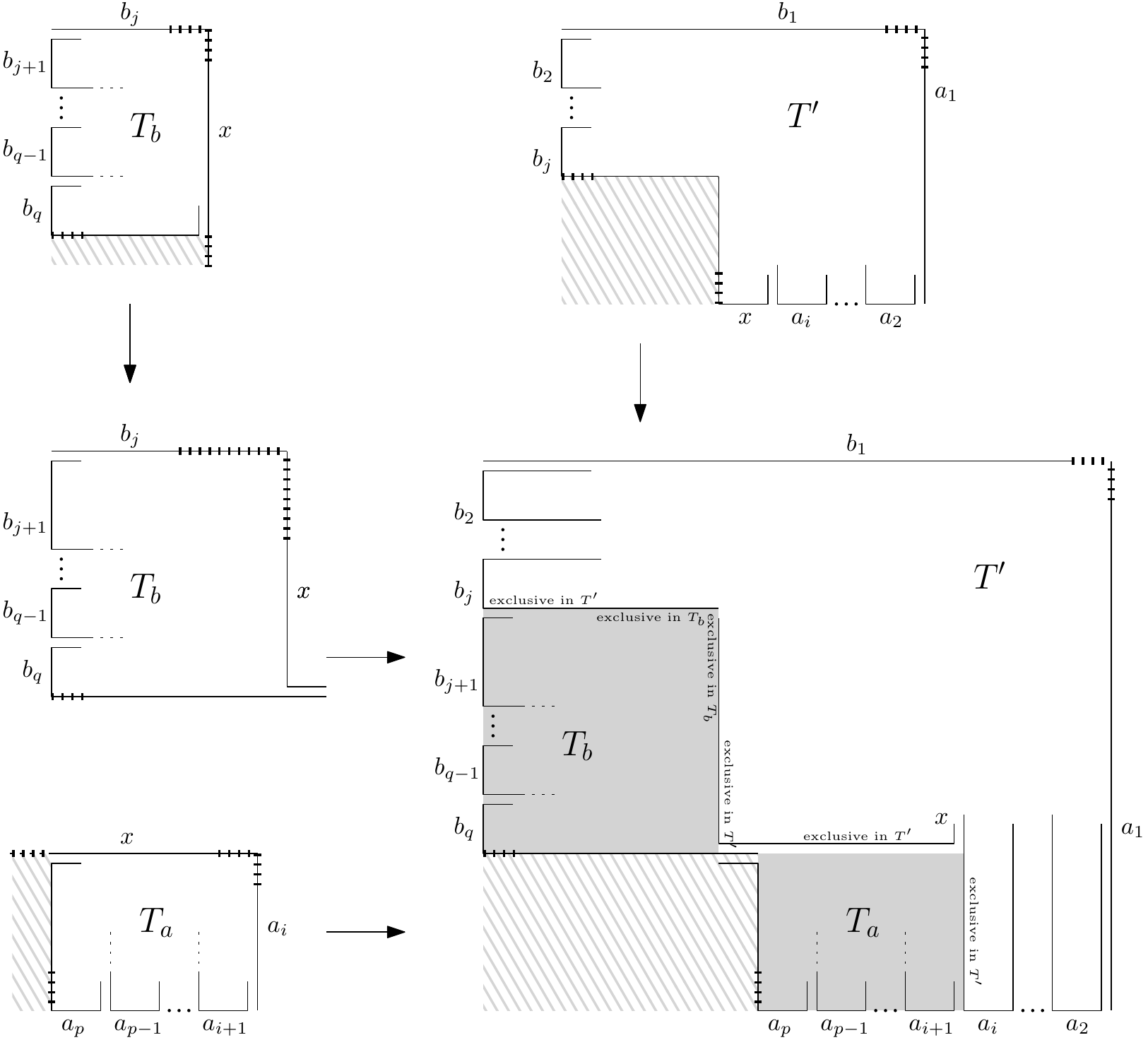}
\caption{Modifying the drawings and then inserting $\Gamma_a$ and $\Gamma_b$ into $\Gamma'$.}
\label{fig:split}
\end{figure}

\paragraph{Splitting}  Assume now that we split $T$, using the common
neighbour $x$ of $a_p$ and $b_q$, into the three subgraphs $T'$, $T_a$ 
and $T_b$.  Let $\Gamma',\Gamma_a,\Gamma_b$ be the recursively 
obtained $B_3$-EPG-representations.  We merge these as illustrated
in Figure~\ref{fig:split}.  Before doing this, we need to modify the
drawings a little bit.

\begin{itemize}
\item
We modify $\Gamma_a$ as follows.
Recall that in $T_a$ vertex $x$ plays the role of $b_1$, and this
is the only vertex on the ``$b$-side''.  Therefore, the point $o_a$
where $\ell(a_p)$ meets $d(x)$ lies on the top side of the bounding
box of $\Gamma_a$, and the only thing on its left is $d(x)$.
We remove $t(a_p)$ from $\Gamma_a$, and replace it by a new $t(a_p)$
that goes leftward from $o_a$.  This represents the same graph by
Invariant~\ref{it:tap}.

\item 
We modify $\Gamma_b$ similarly, but not exactly symmetrically.
Let $o_b$ be the point where $d(b_q)$ meets $\ell(x)$ in $\Gamma_b$.
The only thing below this point is $\ell(x)$, and we cut off this
part.  We also remove $r(b_q)$, and add a grid-edge to the
right of the bottom-right corner that is assigned to both $x$ and $b_q$.
By Invariant~\ref{it:rbq}, this represents the same graph.

\item 
We modify $\Gamma'$ by moving the bottom segments of $a_i,\dots,a_2$
downwards and extending $\ell(a_1)$ downwards.   The amount of extension
is equal to the height of $\Gamma_a$.
\end{itemize}
We also need to stretch all drawings so that they fit within each
other, and furthermore, that segments that belong exclusively to one path
are long enough that we do not have to worry about creating unwanted
shared grid-edges.  
\begin{itemize}
\item Consider the segment $d(b_j)$, which occurs in
both $\Gamma_b$ and $\Gamma'$; we use $d_b(b_j)$ and $d'(b_j)$ to distinguish
these two segments.    
We know that the leftmost grid-edge of $d'(b_j)$ and the rightmost 
grid-edge of $d_b(b_j)$ belongs exclusively to $b_j$.
Expand the leftmost grid-edge of $d'(b_j)$ (by inserting empty columns
into $\Gamma'$) until it is as wide as the non-exclusive rest of $d_b(b_j)$.
Similarly expand the rightmost grid-edge of $d_b(b_j)$ until it is as wide
as the non-exclusive rest of $d'(b_j)$.
\item In a similar way we insert empty rows in $\Gamma_b$ and $\Gamma'$ to ensure
that every grid-edge of $\ell(x)$ that occurs in both $\Gamma_b$ and $\Gamma'$
is exclusive to $x$ in at least one of the drawings.
\item Finally we insert empty columns into $\Gamma'$ to widen $d(x)$ such that
its width equals the width of $\Gamma_a$.
\end{itemize}

Observe that with this, $\Gamma_b$ ``fits'' to the left of $\ell(x)$
and below $d(b_q)$
in $\Gamma'$, and $\Gamma_a$ ``fits'' below $d(x)$ and left of
$\ell(a_i)$ in $\Gamma'$.
Since $x$ and $a_i$ spans the top and right of $\Gamma_a$ and 
$x$ and $b_j$ span the right and top side of $\Gamma_b$,
merging the drawings here means that the two grid-paths of $x,a_i,b_j$
can become one.  
All invariants are easily verified.  Notice that Invariants~(\ref{it:tap}) and
(\ref{it:rbq}) do not need to hold, since in the splitting-step we have $p,q\geq 2$.

We claim that we do not create any unwanted shared grid-edges, i.e.,
grid-edges that belong  to two paths of non-adjacent vertices.  This
could happen only at the boundaries of two merged drawings, i.e., at
$d(b_j),\ell(x),d(x)$ or $\ell(a_i)$.
We do not create any unwanted shared grid-edges at
$d(b_j)$ and $\ell(x)$ due to our modifications.  We do not create
any unwanted shared grid-edges at $d(x)$ since $d(x)$ belongs exclusively
to $x$ in $\Gamma'$ by Invariant~\ref{it:bottom}.  We do not create any unwanted
shared grid-edges at $\ell(a_i)$, since we moved $d(a_i)$ downwards and
hence this stretch of $\ell(a_i)$ is exclusive to $a_i$ in the (modified)
$\Gamma'$.

On the other hand, we must show that for all adjacencies we have created
shared grid-edges.  This is for the most part obvious, since these adjacencies
also exist in one of the subgraphs.  Edge $(b_j,x)$ is not represented
in $\Gamma_b$, but it is represented in $\Gamma'$.  Likewise $(x,a_i)$
is represented (also in $\Gamma'$).  Finally edge $(a_p,b_q)$ is represented
at the rightmost grid-edge of $d(b_q)$, which is shared by $a_p,b_q$ and $x$.

Therefore we found a suitable $B_3$-EPG representation in all cases
and the lemma holds.

\section{Graphs without separating triangles}

We have hence shown that there exists a $B_3$-EPG representation for
all 4-connected triangulated planar graphs.  We now prove 
Corollary~\ref{cor:noSepTriangle}, i.e., we argue that the same result
holds for all planar graphs that have a planar embedding without a separating
triangle, by arguing that they are induced subgraphs of a 4-connected
triangulated planar graph.  

\begin{lemma}
Let $G$ be a planar graph that has an embedding without separating triangle.
Then $G$ is an induced subgraph of a 4-connected triangulation $T$.
\end{lemma}

We first mention that a similar result was proved in \cite{GIP17}, but
they assumed that $G$ has no triangles whatsoever.  One possible proof
of the lemma is hence to take their proof and observe that triangles that
are faces can be handled, simply by not adding anything inside them.  For
completeness' sake, we give here a separate (and possibly simpler)
proof.

\begin{proof}
Let us first assume that $G$ is a {\em wheel-graph}, i.e., it consists of
a cycle $C$ and one vertex $v$ adjacent to all vertices of $C$.  Then the
lemma obviously holds by letting $T$ be the graph obtained by adding a
second vertex $w$ adjacent to all of $C$.  

So we may assume that $G$ is not the wheel-graph.  We then know that we
can add edges to $G$ such that the resulting graph $H$ is a 4-connected
triangulated graph \cite{BKK97}.  Let $H'$ be the graph obtained from
$H$ by subdividing all edges in $H-G$, and observe that $G$ is an induced
subgraph of $H'$.  Furthermore, since $H$ is 4-connected, $H'$ is 2-connected
and the only cutting pairs are those vertex pairs $\{v,w\}$ that are the ends
of an edge of $H-G$.  In particular therefore, all faces of $H'$ are simple
cycles, and no face $f$ of $H'$ can have a {\em chord}, i.e., no two
non-consecutive vertices of $f$ can be adjacent.  For if there were
such a chord $(v,w)$, then $\{v,w\}$ would be a cutting pair, hence in $H$ we
would have had both the edge $(v,w)$ that exists in $H'$ and the edge $(v,w)$
that has been subdivided. This implies a multi-edge in $H$, a contradiction.
Therefore we can now {\em stellate} each face of $H'$, i.e., insert a
new vertex inside each face $f$ and make it adjacent to all vertices of $f$,
without creating a separating triangle.  The graph $T$ that results from
stellating all faces of $H'$ hence is a 4-connected triangulation that
contains $G$ as induced subgraph.
\end{proof}

\bibliographystyle{plain}
\bibliography{refs.bib}

\end{document}